\documentstyle[amssymb,preprint,prb,aps]{revtex}
\tightenlines
\begin{document}
\draft
\title{Evidence for Possible Phase-Separations in 
RuSr$_{2}$(Gd,Ce)$_{2}$Cu$_{2}$O$_{10-\delta}$}
\author{Y. Y. Xue, D. H. Cao, B. Lorenz and C. W. Chu$^{1}$}
\address{Department of Physics and Texas Center for Superconductivity, 
University of Houston,\\
Houston, Texas 77204-5002\\
$^{1}$also at Lawrence Berkeley National Laboratory, 1 Cyclotron Road, 
Berkeley, California 94720;\\
and Hong Kong University of Science and Technology, Hong Kong}
\date{\today}
\maketitle
\begin{abstract}
An unusual thermal-magnetic hysteresis was observed between a minor magnetic
transition around 120~K and the main one at 80~K in superconducting 
RuSr$_{2}$(R,Ce)$_{2}$Cu$_{2}$O$_{10-\delta}$ (Ru1222R) samples, where 
R = Gd or Eu, down to a
submicron length-scale. The observation suggests a possible phase-separation 
and is consistent with the very small but
universal demagnetizing factor observed, which is difficult to reconcile with 
the canted spin-structure previously
proposed. In such a scenario, the unusual superconducting properties of the 
Ru-based cuprates can also be understood naturally.
\end{abstract}

The neutron-powder-diffraction (NPD) observation that the Ru-spins in RuSr$%
_{2}$GdCu$_{2}$O$_{8}$ (Ru1212), which is superconducting below 
20--40~K, are
antiferromagnetically (AFM) aligned below a magnetic transition-temperature 
$T_{M} \approx 133$~K makes its physics more complicated.\cite{lyn,jor}
Although canted Ru-spins have been proposed,\cite{lyn,wil} the
interpretation seems to be debatable. On one hand, the macroscopic
moment of ceramic Ru1212 samples appears as that of a ferromagnet with a
remnant moment of $m_{r} \approx 0.14$ $\mu _{B}$/cell.\cite{ber,wil} On
the other hand, the NPD sets an upper limit of only 0.1 $\mu _{B}$/cell for
the ferromagnetically (FM) aligned spins, $M$.\cite{lyn,wil} Since $M$ should
be larger than $2m_{r}$ in a ceramic,\cite{boz} the two measurements are
difficult to reconcile with each other as canted spins. A possible way to
accommodate these conflicting values is the existance of mesoscopic phase-separations.
Previously, such a possibility has been disregarded based on a $\mu$SR
measurement.\cite{ber} We would like to point out, however, that the
volume-fraction $\approx 2m_{r}/m_{\text{o}}$ of the possible FM
species may be well below the experimental uncertainty of 20\% with the Ru
moment $m_{\text{o}} > 3$ $\mu _{B}$/cell.\cite{ber,but} A similar debate in
manganites further demonstrates that it is difficult to rule out
phase-separations with only a few measurements.\cite{moreo} A re-examination of the
Ru-based cuprates, therefore, is warranted. It should be noted that the
magnetic configuration of Ru1212 is directly related to its
superconductivity (SC). While ``ordinary'' SC may coexist with canted AFM, a
phase-separation will result in the appearance of Josephson-junction arrays
as previously proposed.\cite{chu} The severe granularity of Ru1212 could
thus be understood,\cite{ber,xue,fel} similar to the heterogeneous magnetic
structure of ErRh$_{4}$B$_{4}$.\cite{bul}

As part of our effort, the thermal-magnetic hysteresis of superconducting
RuSr$_{2}$R$_{2-x}$Ce$_{x}$Cu$_{2}$O$_{10-\delta }$ (Ru1222R) with R = Gd or
Eu and $x$ between 0.6 and 0.7, whose magnetic structure is expected to be
similar to that of Ru1212,\cite{fel,bau} is studied. Our data confirm that
there are minor weak-ferromagnetic transitions at $T_{M2} \approx 120$~K and
$T_{M3} \approx 140$~K above the main transition at $T_{M1} \approx 
80$~K,
as reported previously.\cite{fel} A dipole-dipole AFM interaction is further
detected between the $T_{M1}$ and the $T_{M2}$ FM-species with a universal
strength down to a length scale far smaller than the grain size, {\it i.e.}
these two FM-species should coexist in a single structural grain.
Furthermore, the strength, measured as the effective demagnetizing factor of
the dipole field, is too small to be accommodated with a homogeneously
canted-spin structure. A spatial correlation between the two species is
needed. The observation, therefore, suggests a possible phase-separation
between the FM species and a possible AFM matrix. Such a scenario can 
naturally account for many
unusual superconducting properties previously reported in Ru1212/Ru1222.

Ceramic Ru1222 samples were synthesized following the standard
solid-state-reaction procedure. Raw oxide powders were first prepared by
calcination at 400--900 $^{\circ}$C under flowing O$_{2}$. Mixed powder
with a proper cation ratio was then pressed into pellets and sintered at 900 
$^{\circ}$C in air for 24 hr. The final annealing was done at 1090 
$^{\circ}$C for 60 hr after repeatedly sintering at 1000 $^{\circ}$C and
regrinding.\cite{cao} Pulverized samples of Ru1222R with different particle
sizes sorted by the descending speeds in acetone were prepared. The
structure was determined by powder X-ray diffraction (XRD) using a Rigaku
DMAX-IIIB diffractometer. There are no noticeable impurity lines in the
X-ray diffraction pattern within a resolution of a few percent. The grain
sizes of the ceramic samples, the particle sizes of the powders, and the
composition were measured using a JEOL JSM 6400 scanning electron microscope
(SEM) and a JEOL JXA 8600 electron microprobe with attached Wavelength
Dispersive Spectrometers (WDS). The cation-composition is homogeneous within
our experimental resolution ($\approx$ 1--2\%). The magnetizations were
measured using a Quantum Design SQUID magnetometer.

The field-cooled magnetization, $M_{FC}$, at 5~Oe is shown in the inset to
Fig.~\ref{fig1} for a Ru1222Gd powder sample with a particle size of 
$\approx$ 1~$\mu$m 
and diluted with a large amount of epoxy. The particle size was far
smaller than the grain size ($\approx $ 2--20~$\mu$m) of the initial ceramic
sample, and the powder should be regarded as a collection of single grains
with negligible inter-particle (grain) interactions. Two additional
transitions at $T_{M2}$ and $T_{M3}$ can be seen above the main transition
at $T_{M1}$.\cite{note} Below 35~K, the zero-field-cooled magnetization, 
$M_{ZFC}$, becomes negative (not shown) and $M_{FC}$ shows a diamagnetic drop,
indicative of a superconducting transition (inset, Fig.~\ref{fig1}). All of these
observations are in qualitative agreement with the data previously reported. 
\cite{fel}

To determine whether the different transitions are associated with different
structural grains, we examined the thermal-magnetic hysteretic effects; the
spin-alignment of the $T_{M1}$ species in highly diluted powder should not
depend on their history above $T_{M1}$ if one grain contains only one type
of species.\cite{boz} A procedure was thus designed as {\bf A }(field-cooled
under a fixed field of 10~Oe from 200~K to a temperature $T_{S}$) - {\bf B }
(switching field from 10~Oe to a lower field of $H_{S}$ at $T_{S}$) - {\bf C 
}(field-cooled under $H_{S}$ from $T_{S}$ to 60~K) - {\bf D }(raising the
temperature under $H_{S}$ to 200~K) - {\bf E }(field-cooled under $H_{S}$
from 200~K to 60~K) (Fig.~\ref{fig1}). It should be noted that the $H_{S}$ was
exactly the same between steps {\bf C} and {\bf E} and was measured using a
Hall sensor with a resolution of 0.005~Oe. Additional tests that were 
performed
demonstrated that the SQUID magnetometer is suitable to measure
ferromagnets down to $10^{-2}$~Oe if precautions are taken.\cite{xuea}

The $M_{FC}$-jump across $T_{M1}$, $\Delta M_{FC} (H_{S}$,~$T_{S}) 
\equiv M_{FC}\mid _{at\text{ 60 K}} - M_{FC}\mid _{at\text{ 95 K}}$,
and the magnetization just above $T_{M1}$, $M_{FC}(H_{S}$~$T_{S}$)
$\equiv M_{FC}\mid _{at\text{ 95 K}}$, were used here to measure the spin
alignments of the $T_{M1}$ species and the $T_{M2}$ species, respectively.
Both use a unit of emu per cm$^{3}$ Ru1222, {\it i.e.} a ``nominal'' moment
without considerations of the epoxy-filling. During the $T_{M1}$ transition,
the spins of a $T_{M1}$ species should be ordered under both $H_{S}$ and
interaction with the neighboring $T_{M2}$ species. For a dipole-dipole
interaction, the interaction should be proportional to the magnetization of
the $T_{M2}$ species around $T_{M1}$. However, $M_{FC}\mid _{at\text{ 95 K}}$, 
{\it i.e.} the nominal moment at a slightly higher temperature, was used
as an approximation to avoid interferences from the $T_{M1}$ species. A
correlation between $\Delta M_{FC}$ and $M_{FC} (H_{S}$~$T_{S}$) was
indeed observed. The $\Delta M_{FC}$ in step {\bf C}, for example, is
systematically lower than that in step {\bf E}. In particular, the jump 
$\Delta M_{FC}$ in step {\bf C} can even be negative under a positive 
$H_{S}$ up to 0.5--0.6~Oe. It should be noted that this memory effect at low
fields is qualitatively different from the well-known coercive-hysteresis of
ferromagnets.\cite{boz} For example, the observed $\Delta M_{FC}$ at a
fixed $H_{S} = 0.05$~Oe changes sign when $T_{S}$ crosses $T_{M1} \approx 
80$~K (triangles in inset, Fig.~\ref{fig1}). While the positive $\Delta M_{FC}$ at 
$T_{S} < 70$~K represents a typical coercive-hysteresis of the FM $T_{M1}$
domains, the $\Delta M_{FC} < 0$ observed between 80 and 130~K demonstrates
an AFM interaction between the $T_{M1}$ and the $T_{M2}$ domains. This sign
change across $T_{M1}$ demonstrates that the $T_{M1}$ and $T_{M2}$ species
involve different spins and interactions.\cite{note}

This AFM interaction can be seen more clearly in the inset of 
Fig.~\ref{fig2}. The
data of step {\bf E} fall into a straight line through the origin as
expected, while the data of step {\bf C} show a molecular field of 
$-0.6$~Oe.
We interpret the interaction as a demagnetizing field with or without
additional interface interaction.\cite{berk,jac} It is difficult,
unfortunately, to distinguish between these two cases, and we will first
discuss a pure dipole-dipole interaction. In such a case, the demagnetizing
field can be written as $-f\cdot 4\pi M_{FC}(T_{S}$,~$H_{S}$), where $f$
is a constant. Indeed, good linear correlation between the $\Delta M_{FC}$
and $M_{FC}(T_{S}$,~$H_{S}$) was observed in all powder/ceramic samples
examined, supporting the dipole-dipole model (Fig.~\ref{fig2}). An almost universal 
$f = 0.36 \pm 0.1$ was obtained by fitting $\Delta M_{FC} \propto
\lbrack H-f\cdot 4\pi M_{FC}(T_{S}$,~$H_{S}$)] for all cubic-shape
ceramic samples and unaligned powder samples with different Ru1222/epoxy
ratios. This universal $f$ is a surprise since the demagnetizing factor 
$4\pi f$ is expected to be $4\pi f_{\text{o}}V_{R}/(V_{R}+V_{E}$) in the
``nominal'' magnetization-unit used if the $T_{M1}$ and the $T_{M2}$ species
form separated grains, where $f_{\text{o}}$, $V_{R}$, and $V_{E}$ are the
geometric demagnetizing factor and the relative volumes of Ru1222 and epoxy,
respectively. A dilution-independent $f$ demonstrates that the $T_{M1}$ and 
$T_{M2}$ species coexist as nanodomains in the same structural grain, an
indication of possible phase-separations.

It should be noted that geometric corrections are needed to convert 
the $f$ observed in randomly oriented powders to that (identified as 
$g$ in
the discussion below) in a single grain with $H$ along its easy-axis. In
Ru1212, an easy axis has been suggested along the ($a,b$) plane.\cite
{lyn,but} The significant $M$-$H$ hysteresis of both the $T_{M1}$ and $T_{M2}$
species in Ru1222 suggests that a similar situation may exist. Below the
coercive field, the only relevant components of both $H$ and $M_{FC}$ would be
those along the easy axis. In such a case, the intragrain- and
intergrain-contributions to $f$ will be weighted differently if a
correlation exists between the easy axes of the $T_{M1}$ and the $T_{M2}$
species in the same grain. The $f$ of a thin ceramic disk with a
diameter-to-thickness ratio of 5 was measured to verify that. The $f$
observed was 0.25 and 0.5 with $H$ perpendicular and parallel to its axis,
respectively. The values, however, still differ significantly from the
geometric factor of $f_{\text{o}} = 0.125$ and 0.75 expected for a
homogeneous disk, {\it i.e.} the contributions from all neighboring grains
are heavily suppressed. The suppression factor estimated is 2.5 based on
above data, which is in rough agreement with the calculated value of 3
assuming that the intragrain $T_{M1}$ and the $T_{M2}$ species share the
same easy axis but those in adjacent grains are random. It should be
noted that the result is also in agreement with the universal $f$ observed.
The observation, {\it i.e.} two types of magnetic species coexist and share
a common easy axis in the same structural grain, will be another piece of
evidence for the existence of a phase separation.

We estimated the $g$ in a local coordinate system with its $z$-axis being
the common easy axis. The effective field $H^{\ast}$, the $M_{FC}$-jump
across $T_{M1}$, $\Delta M_{FC}^{\ast}(H_{S}$,~$T_{S}$), and the
magnetization just above $T_{M1}$, $M_{FC}^{\ast}(H_{S}$~$T_{S}$),
will be $H\cos\theta_{1}$, $\Delta M_{FC}(H_{S}$,~$T_{S})/\cos\theta 
_{1}$, and $M_{FC}(H_{S}$~$T_{S})/\cos\theta _{1}$,
respectively, where $\theta _{1}$ is the polar angle of $H$. For
unaligned powders, therefore, $g = f/3 \approx 0.12 \pm 0.03$. 
To verify the conclusion that the $g$ is rather small, a Ru1222Eu
sample was partially magnetically aligned by heating a mixture of the powder
and wax to 400~K under a 5~T field. The $M_{FC}\mid _{at\text{ 5 K}}$ of the
cubic sample at 10~Oe were 0.21~emu/$g$ and 0.29~emu/$g$ with $H \parallel $ and 
$\perp $ to its axis, respectively. The $f$ observed, on the other hand,
were 0.37 and 0.29. Assuming an angular distribution of $1-a\cdot\cos
^{2}\theta $, a $g \approx 0.1$ was extrapolated for $a = 0$.

The demagnetizing field of a dipole-array at {\bf r} is the summation of an
average field $-4\pi hM$ and a nearest-neighbor term of $B_{near}$, where $h$
is the demagnetizing factor,\cite{jac} and $h \approx 1/3$ for a
sphere. The $B_{near}$({\bf r}) oscillates rapidly with {\bf r}, but would
be zero after averaging if there is no geometric correlations between {\bf r}
and {\bf r}$_{i}$, where {\bf r}$_{i}$ is the position of the nearest
dipoles. \cite{jac} This can be easily understood since the average field of
a dipole {\bf m} at the origin is $\int [3({\bf m}\cdot {\bf r}){\bf 
r}-{\bf m} \mid {\bf r}\mid^{2}]d^{3}{\bf r}$, which can be factorized as $\propto
\int_{0}^{\pi }(3\cos^{2}\theta -1)\sin\theta d\theta = 0$, where $\theta$ 
is the polar angle of {\bf r}. This average, therefore, will be
only from the dipoles near the sample boundary where the factorization
fails. Their contribution, however, is exactly the $-4\pi hM$ of the
effective surface poles.\cite{jac} In our case, the $h$ should be 
$\approx 1/3$ based on the more-or-less spherical grains observed under SEM.
To confirm that, a sample was measured at different $H$ directions. Strong
anisotropy is expected if the grains are far from spherical. The $g$
deduced, however, is isotropic within 10\%. A significant and positive 
$B_{near} \approx (1/3-g)\cdot 4\pi M = 0.21\cdot 4\pi M$ at the
positions of the FM $T_{M1}$ species, therefore, is required. That, as
discussed above, can be true only if the FM $T_{M1}$ species occupy merely a
small fraction of the samples, {\it i.e.} the nearest $T_{M1}$ and $T_{M2}$
species are aligned along their easy axis. A phase separation between FM
nanodomains and an AFM matrix is suggested.

Additional interface interactions might also exist.\cite{berk} However, the
fact that the $g = 0.12$ observed is almost universal and far smaller
than the geometric factor 1/3 expected requires a sample-independent but
delicate balance between the dipole field and the interface interaction.
This is very unnatural unless both the $T_{M1}$ and $T_{M2}$ nanodomains are
the products of phase-separations.

It might be possible that this phase separation has a chemical origin, 
{\it i.e.} due to the inhomogeneity in Ce doping. In our opinion, however, this
is unlikely: the scenario is not supported by our microprobe data; and the
memory effect is qualitatively the same in our samples with different $x$. To
further probe the topic, the $M_{FC}$ at 1~T (solid diamonds) and the
differential $ac$ $\chi$(T) at 5~T (open circles), as well as the
intercept $M_{\text{o}}$ of the $M$-$H$ loops of a Ru1222Eu sample (Eu was used
to reduce the paramagnetic background) were examined (Fig.~\ref{fig3} and its inset).
It is interesting to note that these parameters, which may serve as a
measure for the volume-fraction of the respective species, vary smoothly
around $T_{M1}$ (Fig.~\ref{fig3}), where a significant anomaly is expected if the 
$T_{M2}$ and $T_{M3}$ species were due to doping inhomogeneity (as
demonstrated by the $M_{FC}$ of Ru1212 at 1--6~T)\cite{wil}. Instead, the
ferromagnetic contribution $M_{\text{o}}$ shows a single transition at 
140~K $\approx T_{M3}>>T_{M1}$ (Fig.~\ref{fig3}), as reported 
before.\cite{fela} The $M_{FC}$ and the 
$ac$ $\chi$(T) show similar behavior. The minor $T_{M2}$ FM species
observed, therefore, is unlikely to be merely due to the doping inhomogeneity.\cite
{note1}

To understand the nature of the $T_{M2}$ nanodomains (no comparable
phenomena observed in Ru1212), a Curie-Weiss (C-W) fit (solid line) was
calculated from $M_{FC}$ at 1~T and above 180~K. A Curie constant of 
$C = 1.03$~emuK/mole ($\approx$ 2.8 $\mu _{B}$/Ru with $\mu $(Ce) $= 2.54$ $\mu
_{B}$ and $\mu $(Eu) $= 0$ $\mu _{B}$) and a C-W temperature $T_{CW}$ of 
84~K $\approx T_{M1}$ were obtained. The increase of the $\chi$ with cooling,
however, significantly slows down below 1.5$T_{M1}$, and is even lower than
the C-W fit below 1.25$T_{M1}$. These are in line with the observed 
$ac$ 
$\chi$(T) with a $dc$ bias of 5~T, which peaks around 120~K $>> T_{M1}$. 
All of these, however, are in great contrast with Ru1212Eu, whose 
$dc$ $\chi$(1~T) stays above the C-W and whose $ac$ $\chi$(T) increases
with cooling down to its AFM transition temperature.\cite{but,xuea} A
short-range AFM correlation, therefore, seems to exist far above $T_{M1}$ in
Ru1222, and may be closely related to the $T_{M2}$ species observed. This is
slightly different from Ru1212, but offers an opportunity to probe the
magnetitic structures below $T_{M1}$.

It should be pointed out that a phase separation of Ru1222 into a FM species
and an AFM species will offer a natural interpretation for many unusual
superconductive properties of Ru1212/Ru1222,\cite{chu,xue} although a 
detailed
structure study is needed to confirm our proposition.

In summary, several magnetic transitions and an unusual thermal-magnetic
memory effect between them were observed in Ru1222. Detailed analysis of the
magnetization under different thermal-magnetic conditions led us to the
suggestion of a phase separation of Ru1222 into FM and AFM nanodomains
inside the crystal grains. Such a suggestion can also account for the
unusual superconducting properties reported in the Ru-based cuprates. A
direct detailed magnetic structure study is warranted to confirm our
proposition.

\acknowledgments
This work was supported in part by NSF Grant No. DMR-9804325, the T.~L.~L. 
Temple Foundation, the John and Rebecca Moores 
Endowment, and the State of Texas through the Texas Center for Superconductivity 
at University of Houston; and at Lawrence Berkeley National Laboratory by 
the Director, 
Office of Energy Research, Office of Basic Sciences, Division of Material 
Science of the U.~S.~Department of Energy under Contract No.~DE-AC0376SF00098.

\begin{figure}[tbp]
\caption{$M(T)$ in a procedure discussed in text. From top to bottom: $H_{S}
= 1.4$, 1.02, 0.63, $-0.04$, $-0.22$, and $-0.6$~Oe. Solid symbols: data in steps
{\bf B}-{\bf C}; open symbols: data in the section {\bf D}-{\bf E}. Inset: $\bigcirc$ - $M_{FC}$
under 5~Oe; the four vertical arrows show $T_{c}$, $T_{m1}$, $T_{m2}$, 
and $T_{m3}$, respectively; $\blacktriangle$ - $\Delta M_{FC}$ at $H_{S} = 
0.05$~Oe but with different $T_{S}$.}
\label{fig1}
\end{figure}

\begin{figure}[tbp]
\caption{$\Delta M_{FC}= M_{FC}$(95~K)$-M_{FC}$(60~K) against
the effective field $H-f\cdot 4\pi M$(105~K, $H_{S}$) 
($f = 0.36$ in this sample). Inset: $\Delta M_{FC}$ {\it vs.} $H$. Open
symbols: data from steps {\bf D}-{\bf E}. Solid symbols: the data from 
steps {\bf B}-{\bf C}.}
\label{fig2}
\end{figure}

\begin{figure}[tbp]
\caption{The susceptibility $\chi '$ of a Ru1222Eu sample. 
$\diamondsuit$: from $M_{FC}$ at 1~T; thin solid line: Curie-Weiss fit based
on the $M_{FC}$ data between 180 and 350~K; $\bigcirc$: from $ac$
susceptibility with a $dc$ bias of 5~T; -$\bigtriangleup$-: the
ferromagnetic contribution represented as the $M_{\text{o}}$ of the inset. Inset:
the average $M_{ave} = (M_{inc} + M_{dec})/2$, where $M_{inc}$ and 
$M_{dec}$ are the magnetization in the $H$-increase branch and $H$-decrease 
branch of a $\pm 5$~T $M$-$H$ loop, respectively. The curves were measured at 
60, 70, 80, 90, 100,
110, 120, 130, and 140~K (from top to bottom).}
\label{fig3}
\end{figure}
%
%

%
%


\begin{references}
\bibitem{note}  The effect of the $T_{M3}$ species is rather small and
cannot be unambiguously identified within our resolution.
\bibitem{note1}  It should be noted that the small-but-universal $g$
will still require a phase separation between an AFM matrix and FM
nanodomains even if the $T_{M2}$ species were impurities.
\end{references}
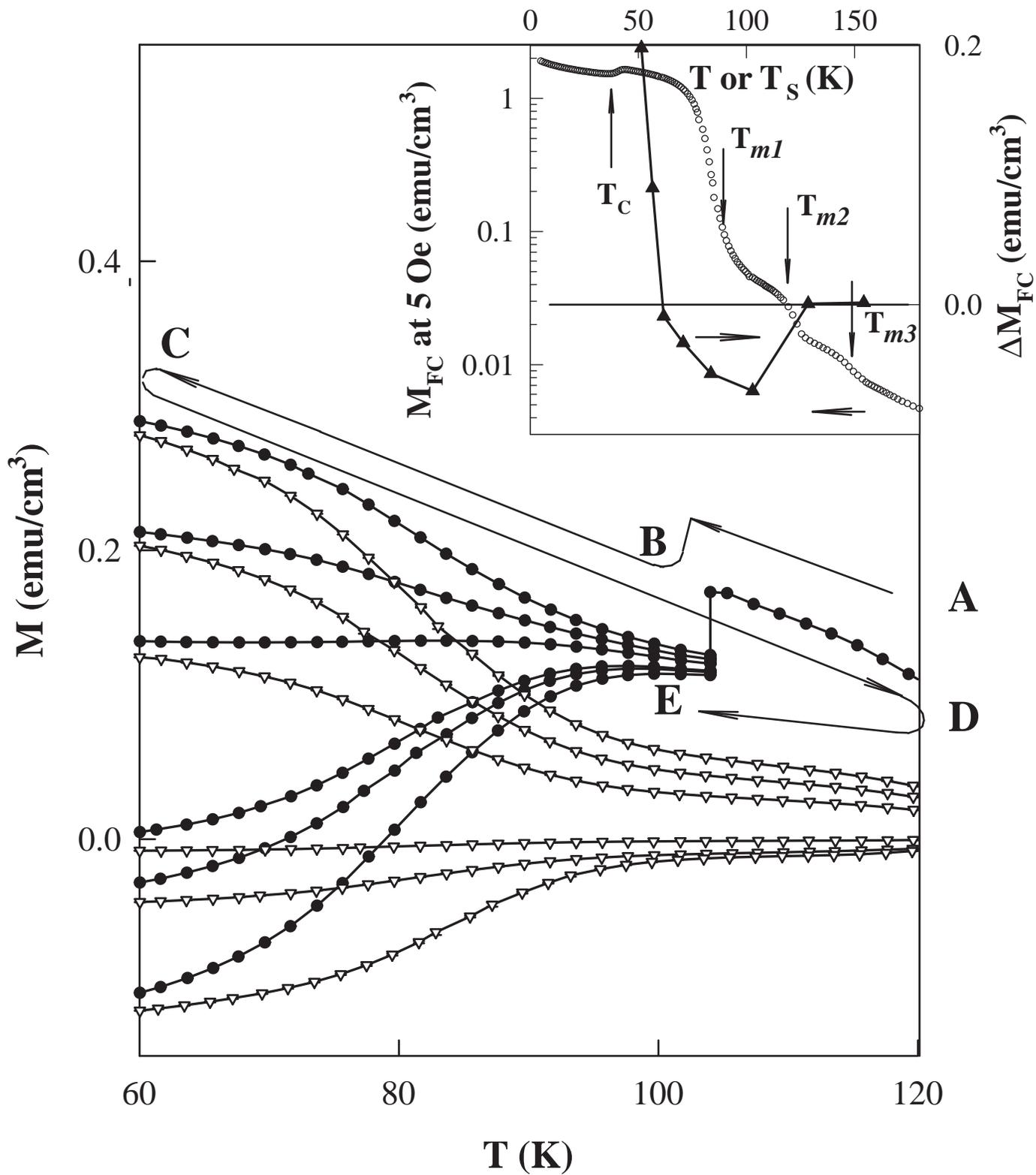
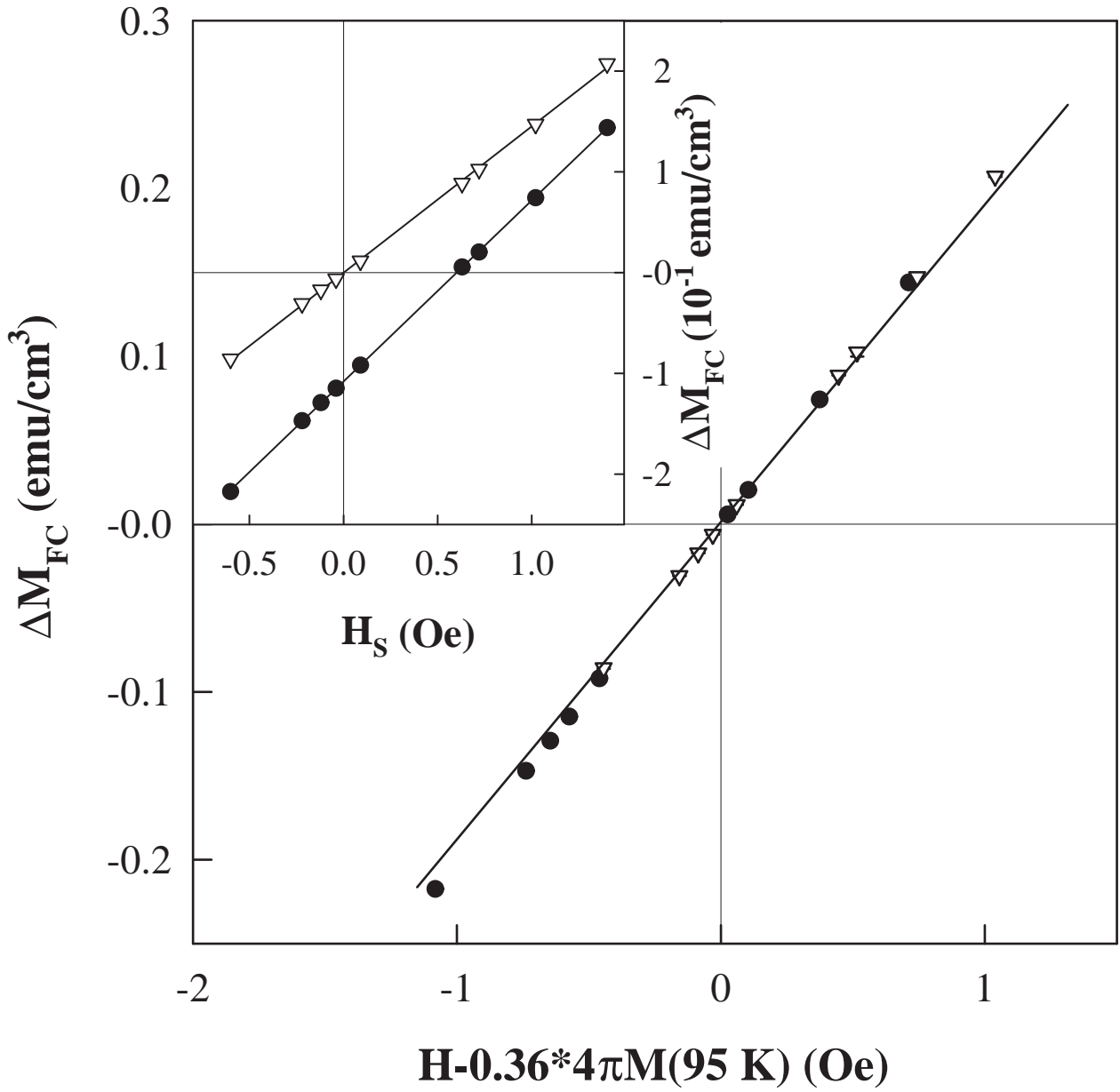
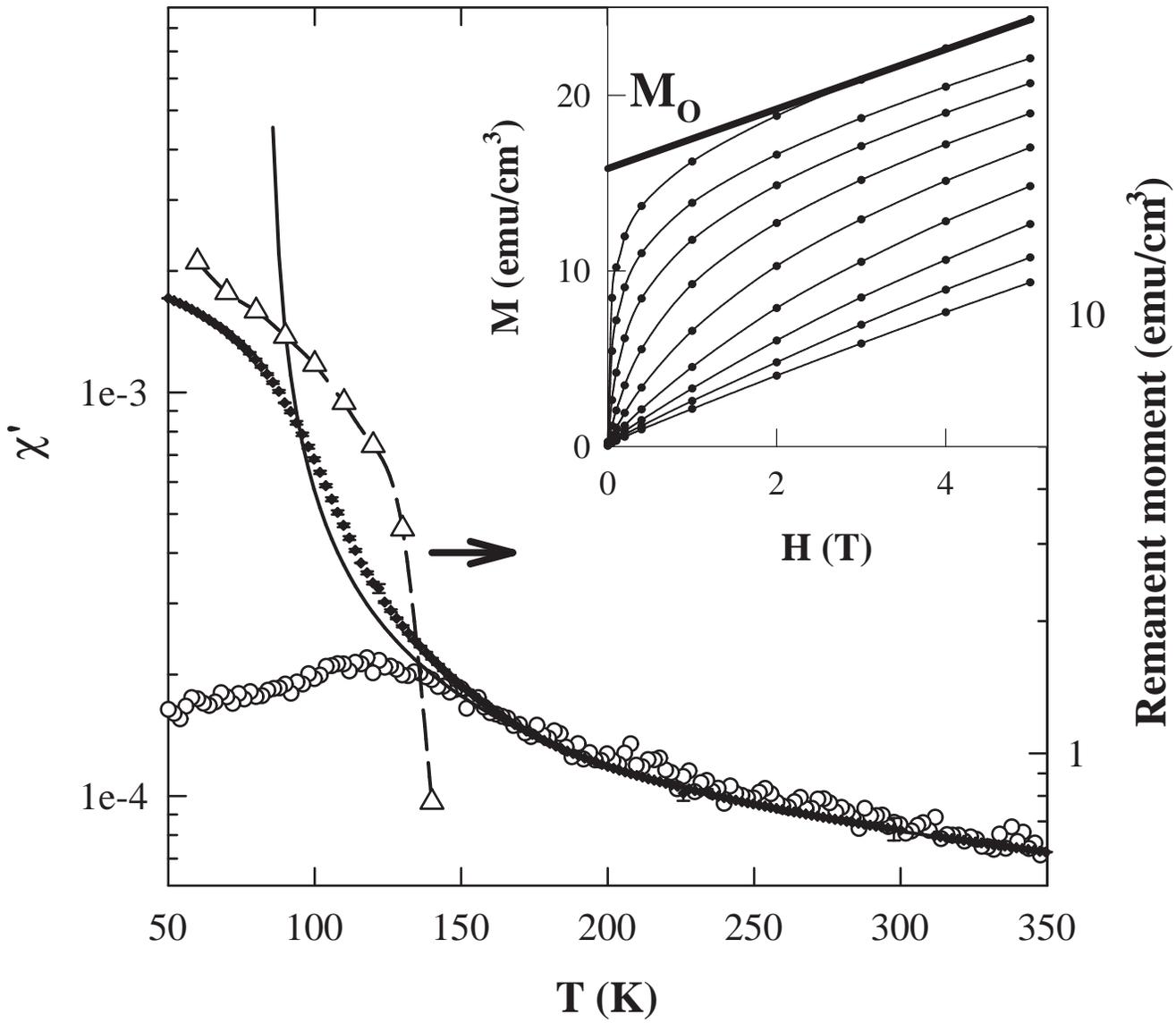
\end{document}